\documentclass{article}
\usepackage{spconf,amsmath,graphicx,amsfonts,hyperref,dsfont}
\usepackage{xcolor}
\usepackage{booktabs}
\usepackage{arydshln}

\title{On the role of Visual Context in Enriching Music Representations}

\name{Kleanthis Avramidis\qquad Shanti Stewart\qquad Shrikanth Narayanan}
\address{Signal Analysis and Interpretation Lab, University of Southern California, Los Angeles, CA 90089}

\begin{document}
\ninept
\maketitle

\begin{abstract}
Human perception and experience of music is highly context-dependent. Contextual variability contributes to differences in how we interpret and interact with music, challenging the design of robust models for information retrieval.  Incorporating multimodal context from diverse sources provides a promising approach toward modeling this variability. Music presented in media such as movies and music videos provide rich multimodal context that modulates underlying human experiences. However, such context modeling is underexplored, as it requires large amounts of multimodal data along with relevant annotations. Self-supervised learning can help address these challenges by automatically extracting rich, high-level correspondences between different modalities, hence alleviating the need for fine-grained annotations at scale. In this study, we propose VCMR -- \textit{Video-Conditioned Music Representations}, a contrastive learning framework that learns music representations from audio and the accompanying music videos. The contextual visual information enhances representations of music audio, as evaluated on the downstream task of music tagging. Experimental results show that the proposed framework can contribute additive robustness to audio representations and indicates to what extent musical elements are affected or determined by visual context. \footnote{Code and results available at https://github.com/klean2050/VCMR}
\end{abstract}

\begin{keywords}
Self-Supervised Learning, Multimodal Learning, Visual Context, Music Information Retrieval, Music Tagging
\end{keywords}

\begin{figure*}[t!]
    \centering
    \includegraphics[scale=0.28]{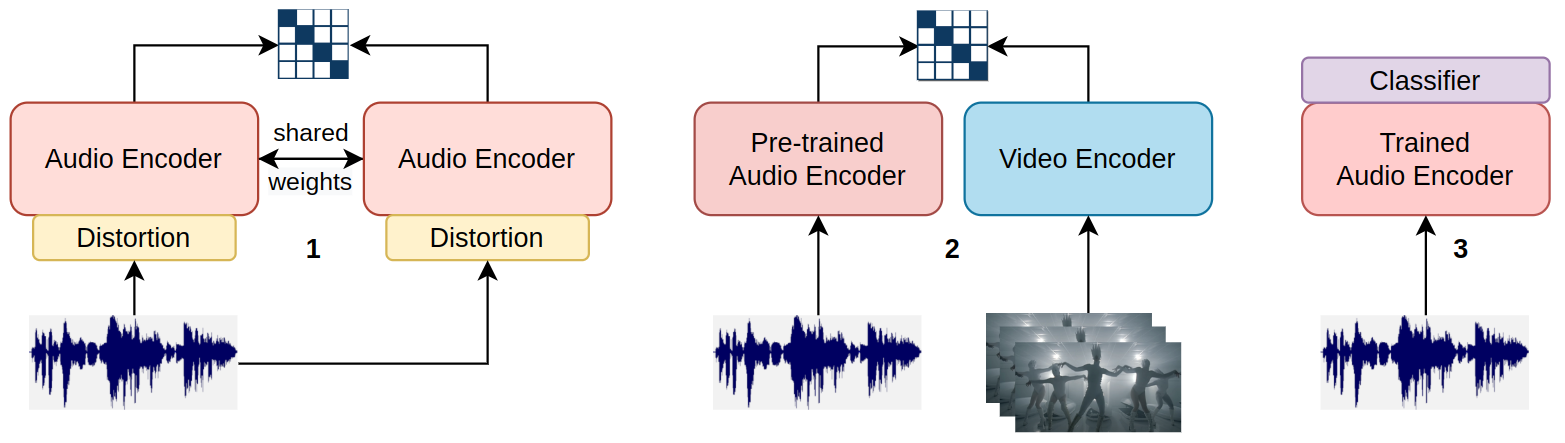}
    \vspace{-0.25cm}
    \caption{Overview of the proposed VCMR framework. 1) Contrastive pre-training of music audio 2) Multimodal contrastive pre-training between audio and aligned video frames 3) Supervised fine-tuning on music auto-tagging.}
    \vspace{-0.25cm}
    \label{fig:my_label}
\end{figure*}

\section{Introduction}
\label{sec:intro}
Music is one of the most popular forms of art and entertainment and is known to encapsulate strong affective characteristics. Music plays a central role in human personal and social experiences (e.g., celebration, grief, nostalgia, or stress). Questions on why music is so intimately connected to human experiences are predominantly related to its ability to induce and regulate our feelings \cite{jakubowski2021, sacks}. From a computational point of view however, we are still far from fully understanding the mechanisms of emotion induction, both regarding the musical structure that causes it and the human physiology and psychology that is associated with it \cite{thompson2022psychological}.

Music Information Retrieval (MIR) has been advancing rapidly, especially after the proliferation of machine learning, in finding markers in music that are associated with pitch, tempo, genre and other informative tags \cite{won2020evaluation}. However, a large number of musical features is still only weakly modeled from audio information. For instance, multiple co-playing instruments make their identification harder. Similarly, discerning mood-related information in music is inherently subjective especially when specific context is not available. To overcome this challenge, research in computational music perception has underscored the importance of leveraging contextual information \cite{barrett2011context} from multiple sources. While there have been efforts in leveraging multimodal information for music analysis---such as lyrics and chords \cite{greer2019learning}---in this study we focus on musical context related to the social or personal circumstances that are associated with the presentation of the music pieces. Music videos, especially professionally-rendered official releases, capture such elements, since they are curated for the purpose of inducing specific responses in the audience. For example, music in films accompanies visual scenes to match and intensify the theme of the scene, whereas a music video of a party song will depict the underlying mood~\cite{BenMa-journal.pone.0249957}.

Recent advances in machine learning have paved the way to leverage extensive amounts of rich multimodal information without the need for explicit annotations. Self-supervised learning (SSL) is an approach that aims to leverage huge quantities of available unlabeled data to enhance unsupervised training and out-perform fully supervised techniques \cite{chen2020big, henaff2020data}. Among the different learning objectives that have been proposed, contrastive learning methods \cite{chen2020simple} are among the most popular and directly applicable to multimodal learning problems \cite{guzhov2022audioclip, alwassel2020self, alayrac2020self}. While SSL is still not fully tapped within the MIR community, the available media sources are abundant. Notably, YouTube has been a popular repository for music creators, and we utilize it in this work for retrieving official music videos.

In this paper, we introduce Video-Conditioned Music Representations (VCMR), a self-supervised multimodal framework that leverages contextual information from music videos to enhance audio music representations. The proposed model takes inspiration from similar works in audio representation learning, while conditioning the learned features with contextual information from audio-video pairings. Our experiments show that the multimodal model achieves improved performance in the downstream task of music tagging, relative to the equivalent model pre-trained only on audio. VCMR further demonstrates robustness across different training settings, input resolutions and cases of labeled data scarcity.
\vspace{-0.1cm}

\section{Related Work}
\label{sec:related}
\vspace{-0.1cm}

Our approach falls in the category of multimodal representation learning algorithms. Representation learning identifies features in data streams that are both efficient and robust in explaining and predicting a number of downstream tasks or labels. These representations are typically learned from a large pre-training dataset in a supervised or unsupervised manner. Thereafter, latent model activations are extracted and used at a specialized downstream task.


Self-supervised learning is a recently proposed variation of unsupervised representation learning that leverages training objectives derived solely from unlabeled data. Our study is influenced by SimCLR \cite{chen2020simple} and CLIP frameworks \cite{radford2021learning}. SimCLR is an efficient SSL model in the vision domain, but has also been effective in speech and audio understanding \cite{pascual2019learning,saeed2021contrastive}. By forcing representations of perturbed versions of the input to be close in the embedding space, SimCLR effectively identifies semantically important information in the data. On the other hand, CLIP is trained on a wide variety of images with language supervision, retrieved from the web. In a similar fashion, both an image encoder and a text encoder are trained to identify matching image-text pairs. CLIP-based learning has already shown robust performance in many audio-related \cite{wu2022wav2clip}, multimodal tasks \cite{guzhov2022audioclip, zellers2021merlot}, but its application to music audio is still limited.

SSL applied in Music Information Retrieval, while relatively less explored, holds great potential for music tagging and recommendation systems. CLMR \cite{spijkervet2021contrastive} adapts the SimCLR framework to the music audio domain. By applying random 1D transformations like pitch shifting, noise, delay and reverb, the authors manage to learn rich music representations, applicable to music auto-tagging.
Regarding multimodal learning between audio and video modalities, most related works focus on cross-modal retrieval for recommendation purposes \cite{suris2022s,thao4189323emomv}. However, these studies typically process unconstrained videos that vary in quality and underlying semantics, while rarely considering the impact of visual context on specific musical cues. Modeling music along with language has also gained attention recently \cite{huang2022mulan,manco2022contrastive}, due to the advances in language models. \vspace{-0.1cm}

\section{The VCMR Framework}
\label{sec:method}

Given a dataset of music videos, consisting of raw audio and video data, we design a self-supervised multimodal framework to enrich music audio representations by conditioning them on visual context. The framework consists of a sequence of 3 stages: 1) contrastive pre-training of music audio 2) multimodal contrastive pre-training between audio and aligned video frames 3) supervised fine-tuning of the learned music representations on music auto-tagging. \vspace{-0.2cm}

\subsection{Input Audio \& Video features}

We use raw audio data at 16 kHz. For the video modality, we use the pre-trained CLIP~\cite{radford2021learning} model to generate robust video features for multimodal learning. We first downsample each video to 5 frames per second, use CLIP to produce a 512-dimensional feature vector for each frame, and then average these vectors every second. Thus, each second of video is represented by a single 512-dimensional feature vector. It has been shown \cite{suris2022s} that the CLIP model produces robust representations for videos since it is trained on a larger corpus of image--text pairs than models trained on ImageNet. \vspace{-0.2cm}

\subsection{Music Pre-Training}

For the first step of our algorithm we perform self-supervised pre-training on the single modality of music audio waveforms. Our model takes an input music waveform, extracts two random sub-segments of 6.15 seconds and produces two augmented views by applying a series of transformations in a randomized manner (e.g., pitch shifting, gaussian noise, reverb and frequency filtering). The resulting 2 views of the input are passed through a SampleCNN \cite{lee2017sample} that has been efficient in encoding music signals \cite{choi2017tutorial,won2020evaluation}. The input resolution is determined by the architecture of SampleCNN, since its layer configuration is built so as to transform an input waveform to a single feature over 512 filter channels. This limits our design choices to specific combinations of encoder blocks and kernel sizes. A resolution of 6.15 seconds is achieved by only altering the kernel size of SampleCNN's first convolutional layer. We chose the specific resolution to account for more available video context during training, while we also present an ablation study on the specific parameter.

The encoder outputs two 512-D embeddings, one for each view. To facilitate the contrastive learning approach, the two views are first projected to an embedding space of lower dimension and then contrasted within the training batch, where positive pairs are formed between the two views of the same input and all the remaining combinations of views are considered negative pairs. The resulting objective is the NT-Xent loss, adopted from the SimCLR study:
\begin{equation}
\ell_{i, j}=-\log \frac{\exp \left(\operatorname{sim}\left(z_i, z_j\right) / \tau\right)}{\sum_{k=1}^{2 N} \mathds{1}_{[k \neq i]} \exp \left(\operatorname{sim}\left(z_i, z_k\right) / \tau\right)}
\end{equation}
Here, $N$ is the batch size and $(z_i, z_j)$ the considered pair of views. We utilize the cosine similarity $\text{sim}(.)$ as the pairwise distance metric between the embeddings and we empirically tune the temperature parameter $\tau$. After training, we discard the latent projector that we used to compute the NT-Xent loss.

\subsection{Multimodal Pre-Training}

For the second step of our algorithm we perform multimodal contrastive pre-training between the audio and the video modality. Specifically, we follow a similar approach to the first step, except that we now form multimodal pairs and the encoders differ. For the music modality we retrieve the previously trained SampleCNN backbone, whereas for the video modality we use a simple 2-layer LSTM architecture on the pre-trained embeddings. We follow this approach since we are primarily interested in \textit{conditioning} the audio embeddings with the respective visual context, instead of learning a bidirectional common representation. At this stage we do not apply any augmentation to the music modality other than the random input cropping. For the video modality we likewise select the segment that corresponds to the extracted audio interval.

Similar to the previous step, we get a 512-D output embedding for the music modality and a 512-D embedding for the video modality through a single fully-connected layer. We then project the embeddings to a lower dimensional space and apply an NT-Xent objective within the training batch. Positive pairs are now considered only the music-video pairs of the same video. After training we discard both projectors and video LSTM network to solely store the music audio backbone for the fine-tuning tasks. \vspace{-0.1cm}

\subsection{Fine-Tuning Framework}

The evaluation of the learned representations is commonly done through a series of downstream tasks, in which the backbone model is kept frozen and its output representations are tested after applying a single or multi-layer perceptron (MLP). Here we use the pre-trained music encoder as a backbone and apply a 2-layer MLP to transform the 512-D embeddings to the label space of our tasks. \vspace{-0.1cm}

\section{Experimental Setup}

\subsection{Music Video Dataset}

To pre-train our model (first and second step) we need a large-scale dataset of music videos that provides both music and video modalities. However, due to inherent limitations in collecting music-related data, such as copyright issues, carefully curated datasets of this kind are scarce and of limited scale, e.g., Harmonix \cite{nieto2019harmonix}. In this study, we obtain official music videos directly from YouTube. YouTube-music-video-5M is a list of such clips, provided as YouTube identifiers and released in 2018 by K. Choi\footnote{https://github.com/keunwoochoi/
YouTube-music-video-5M}. Among the listed 5,119,955 videos we extract 20150 (about 700 hours) for our pre-training task, also providing their respective YouTube IDs.

Initially, more than 100k tracks were selected at random and downloaded from the first 2 lists of YouTube-music-video-5M. After manually inspecting a subset of the videos, we realized that a significant number of them (about 20\%) were not official music videos but either lyric videos, audio releases of still images or even song covers of amateur musicians. We thus proceeded to clean the dataset and remove all such files. A discriminative characteristic of these, compared to official music videos, is that they typically include very sparse scene changes. Thus, we extracted scene information with the PySceneDetect tool\footnote{https://github.com/Breakthrough/PySceneDetect}  and automatically discarded all videos that were shown to include scenes lasting more than 30 seconds. \vspace{-0.1cm}

\subsection{Evaluation Protocol}

We consider music tagging as the downstream task to evaluate our approach and observe to what extent visual context enriches musical information. Following the literature, we use average area under the receiver operating characteristic curve (ROC-AUC) and average precision (PR-AUC) to measure model performance. PR-AUC is considered because ROC-AUC can be over-optimistic for imbalanced cases \cite{davis2006relationship}. We evaluate on the test set of the utilized datasets, averaged over three training sessions with different random seeds. We divide each test track into segments of the selected length with 50\% overlap and average their predictions. We list the datasets below:

\textbf{MagnaTagATune} (MTAT) dataset \cite{mtat} consists of 25,000 music clips from 6,622 unique tracks. These are provided in pieces of about 30 seconds each, of which we segment fragments for training, using the pre-configured training splits. We test our model in predicting the top 50 semantic tags in this dataset. These are provided by human listeners and describe both analytical elements like instrumentation, and affective like mood or theme.

\textbf{MTG-Jamendo} (MTG-M) \cite{bogdanov2019mtg} is an open-source dataset for music auto-tagging. It contains over 55,000 full audio tracks with 195 tags categories (87 genre tags, 40 instrument tags, and 57 mood and theme tags). It is built using music available at Jamendo under Creative Commons licenses and tags provided by content uploaders \footnote{https://jamendo.com}. For our study we use a subset of the dataset that has been used in the Emotion and Theme Recognition in Music Task within the MediaEval challenge \cite{mediaeval}. This \textit{autotagging-moodtheme} subset includes 18,486 audio tracks with mood and theme annotations. In total, there are 57 tags, distributed in a multi-label fashion. \vspace{-0.1cm}

\subsection{Implementation Details}

The SampleCNN encoder takes as input waveforms of 6.15 seconds, randomly chunked from the 15-second audio input, at a sample rate of 16 kHz. It consists of 9 consecutive 1-D convolution blocks with 3-sample kernels. Each convolution layer is followed by batch normalization, ReLU activation and max pooling layers, while the output flattened embedding size is 512 samples. The model yields a lightweight scheme of 2.6M parameters. For details regarding the augmentation transforms applied to the audio waveforms, the reader is encouraged to consult CLMR \cite{spijkervet2021contrastive}.

We used a batch size of 128 samples to pre-train and fine-tune our models (64 during music pre-training due to memory constraints). We pre-trained for 50 epochs, determined empirically from the validation loss history. We then fine-tuned for up to 50 epochs on each downstream task. In all experiments we used Adam optimizer. During multimodal pre-training we kept the first 4 encoder blocks frozen, while, during fine-tuning, the whole encoder was frozen. We used a global learning rate of 0.001, with a weight decay of 1e-6. The temperature parameter was empirically determined as $\tau = 0.1$. Our primary models were trained on 4 $\times$ NVIDIA GeForce RTX 2080 GPUs for a total of 17 hours end-to-end.

\begin{table}
\centering
  \label{tab:results_main}
  \begin{tabular}{lcccc}
    \toprule
    \textbf{Model}&\textbf{Dataset}&\textbf{ROC-AUC}&\textbf{PR-AUC}\\
    \midrule
    Audio-Only (ours) & MTAT& 77.4 \% & 22.6 \% \\
    \textbf{VCMR} (ours) & MTAT& \textbf{89.1} \% & \textbf{35.3} \% \\
   $^{*}$ CLMR \cite{spijkervet2021contrastive} & MTAT & 87.8 \% & 33.1 \% \\
    $^{\dagger}$ SampleCNN \cite{samplecnn} & MTAT & 90.6 \% & 44.2 \% \\
    $^{\dagger}$ HarmonicCNN \cite{harmonic} & MTAT& 91.3 \% & 46.1 \% \\
    \midrule
    Audio-Only (ours) & MTG-M & 57.8 \% & 5.3 \% \\
    \textbf{VCMR} (ours) & MTG-M & \textbf{70.7} \% & \textbf{10.1} \% \\
   $^{\dagger}$ MediaEval 2020 \cite{knox2020mediaeval} & MTG-M & 76.6 \% & 15.0 \% \\
  \bottomrule
\end{tabular}
\vspace{-0.1cm}
\caption{Downstream music tagging performance on the MagnaTagATune (MTAT) and MTG-Jamendo (MTG-) datasets, compared to CLMR \cite{spijkervet2021contrastive} and fully supervised models. $^{*}$For a fair comparison, we report the transfer results of CLMR pre-trained on Million Song Dataset \cite{msd}. $^{\dagger}$We include fully supervised models as reference.}\vspace{-0.4cm}
\end{table}

\section{Experimental Results} \label{sec:exp}

\subsection{Music Tagging Benchmark}

The main aim of this study is to evaluate the quantitative and qualitative effects of conditioning audio representation learning of music to visual media context. In Table~\ref{tab:results_main} we present the performance scores on the music tagging downstream task for the two datasets. The proposed model (VCMR) significantly out-performs its audio-only counterpart in all cases. For MagnaTagATune, VCMR scores 11.7\% higher in ROC-AUC and 12.7\% higher in the more challenging PR-AUC score. For MTG-Jamendo, VCMR is similarly 12.9\% and 4.8\% better, respectively, underscoring the additional information related to mood and affect. Our implementation is also better than CLMR, reported in a transfer learning setting, similar to our experimental setup. VCMR still underperforms popular supervised baselines, also listed in Table~\ref{tab:results_main}, however this comparison is essentially bound to the scale of our pre-training dataset. \vspace{-0.2cm}

\subsection{Tag-wise Performance}

\begin{figure}[t!]
    \centering
    \includegraphics[scale=0.365]{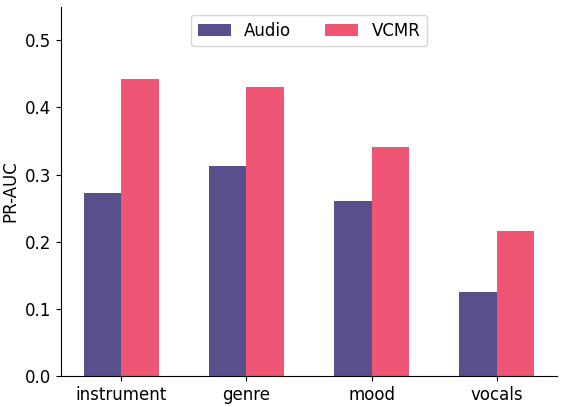}
    \vspace{-0.25cm}
    \caption{Average tag-wise performance on MagnaTagATune, after grouping the provided set of tags into 4 semantic categories.}
    \vspace{-0.3cm}
    \label{fig:tags}
\end{figure}

The derived results suggest that visual conditioning enhances music representations in media in multiple ways. However, we are also interested in determining which music elements are mainly affected and to what extent. To this end, we further look into tag-wise performance differences between VCMR and the audio-only baseline. For MTG-Jamendo, the utilized subset is already specialized in affective elements, hence we focus on MagnaTagATune, whose labels we group into 4 semantic classes: Genre, Mood, Instruments and Vocals. In Figure~\ref{fig:tags} we compare the average performance of the two models in each of these groups. We deduce that VCMR is consistently better in all categories, and provides the largest improvements for the instrument and vocals categories. Intuitively, visual depictions of the instruments played during a song, or who is singing, assists the network in disentagling these tags. Similar trends occur when evaluating the framework based on PR-AUC scores. \vspace{-0.2cm}

\subsection{Input Resolution}

Our base network implementation uses audio and video data at samples of 6.15 seconds. This resolution is determined by the architecture choices of the SampleCNN encoder and was chosen to account for larger context in the learning process, especially with respect to video conditioning. We evaluate this premise by running an ablation study over alternative input resolutions that are permitted by the SampleCNN structure. Specifically, we consider input lengths of 3.07, 3.69 and 4.96 seconds, apart from our base implementation. We note that \cite{spijkervet2021contrastive} used 3.69 seconds input length, whereas \cite{knox2020mediaeval} used 5 seconds, but within a different architecture. 

\begin{figure}[t!]
    \centering
    \includegraphics[scale=0.39]{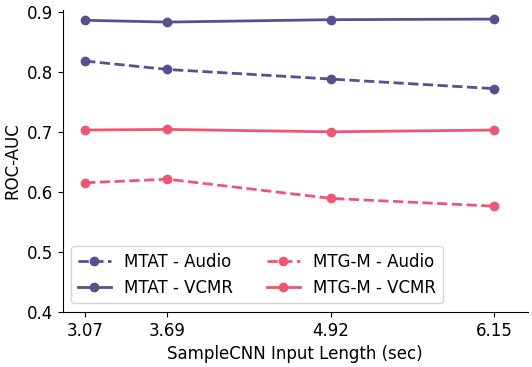}
    \vspace{-0.25cm}
    \caption{Ablation Study on input resolution. The proposed model is trained on input lengths in the range of 3.5 to 6.5 sec. The exact length is determined by the SampleCNN architectural constraints.}
    \label{fig:input_res}
    \vspace{-0.2cm}
\end{figure}

We plot the results for both datasets in Figure~\ref{fig:input_res}. VCMR appears robust across time-scales, with the increased resolution resulting in a slightly increasing trend. On the other hand, the audio-only model shows a reverse trend, where smaller input sizes account favorably for its performance. The discrepancy of the optimal input length between the two models indicates that audio learning focuses on learning small-scale features, whereas video conditioning can disseminate information and context from multiple scales. Still, in all time scales, VCMR out-performs the audio-only baseline.\vspace{-0.2cm}

\subsection{Data Efficiency}

\begin{figure}[t!]
    \centering
    \includegraphics[scale=0.39]{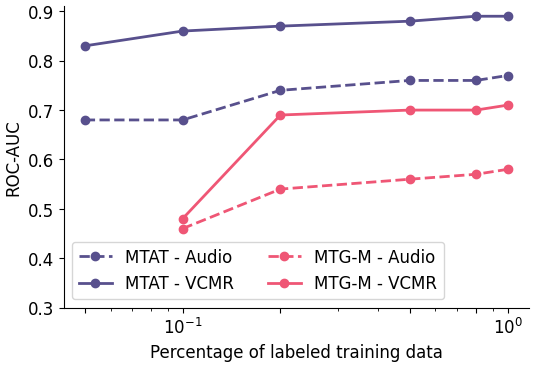}
    \vspace{-0.2cm}
    \caption{Simulation of data scarcity in music tagging. The proposed pre-trained model is fine-tuned on \{5, 10, 20, 50, 80\}\% of the available training data while being tested on the entire test set.}
    \label{fig:scarcity}
    \vspace{0.1cm}
\end{figure}

An essential component of an efficient learning representation is the robustness it shows when fine-tuned with limited data sources (data scarcity). For the task of music tagging, we simulate data scarcity by constraining the available labeled training samples for both datasets. The results, plotted in Figure~\ref{fig:scarcity}, show a slight increasing trend, as expected in all cases. VCMR indeed shows smaller incremental improvements as data size increases, almost matching full training performance with 10\% of the training data. This fact indicates its additive robustness compared to the audio-only model.

\section{Conclusion}
\label{sec:conc}

In this study we proposed VCMR, a multimodal framework for music representations that we train on music audio, conditioned on accompanying visual context, obtainable from official video releases. VCMR enhances music audio representations without embedding any explicit visual features, as evaluated on music auto-tagging. It particularly showed improved robustness both in terms of the utilized training data and the input resolution, compared to the audio-only baseline. In the future, we will incorporate additional objectives, like cross-modal retrieval, to evaluate video conditioning, whereas a direction of interest would be the investigation of which explicit visual features contribute to the model's improved performance.

\vfill\pagebreak
\bibliographystyle{IEEEbib}
\bibliography{refs}

\end{document}